\documentstyle[12pt]{article}
\normalbaselines
\begin{document}

\begin{center}
{\large \bf SYMMETRIES OF THE FREE SCHR\"ODINGER EQUATION\\[2mm]}
G.A. Kotel'nikov\\{\it RRC "Kurchatov Institute",
123182 Moscow, Russia}\\
(e-mail: kga@kga.kiae.su)\\[3mm]
\end{center}

\begin{abstract}
    An algorithm is proposed for research into the symmetrical properties 
of theoretical and mathematical physics equations. The application of this 
algorithm to the free Schr\"odinger equation permited us to establish that 
in addition to the known Galilei symmetry the free Schr\"odinger equation 
possesses also the relativistic symmetry in some generalized sense.  This 
property of the free Schr\"odinger  equation  permits  the equation to be 
extended  into  the  relativistic  area  of movements of a particle being 
studied.
\end{abstract}

\section{Introduction}
\label{1}
The symmetry properties of theoretical and mathematical physics
equations contain the important information about objects of research.
The relativization of space-time obtained by studying the space-time 
symmetries of Maxwell equations may be an example.
\par A number of approaches was proposed for studying symmetries: the
method of replacing the variables \cite{Voi87}, \cite{Gel58};
the classical Lie algorithm \cite{Ibr83}, \cite{Olv89}; the modified Lie
algorithm \cite{Nie72}, \cite{Fus90}; the algorithmes of finding 
the generelized and non-Lie symmetries  \cite{Ibr83}, \cite{Olv89},
\cite{Fus83}, \cite{Zhe86}; the theoretical-algebraic approach 
\cite{Mal79}, \cite{Bar80}, \cite{Lez86};
the method for studying the conditional symmetries \cite{Olv86}, 
\cite{Vor86}, \cite{Fus89}; the renormgroup concept \cite{Shi94},
\cite{Kov96}.
The purpose of the present work is formulation of the algorithm \cite{Kot93},
enabling to obtain additional information. The elements of this algorithm  
were elaborated by studying symmetries of D'Alembert and Schr\"odinger 
equations \cite{Kot83}, \cite{Kot83'}, as well as by proving the Galilei
symmetry of Maxwell equations \cite{Kot86}, \cite{Kot91}, \cite{Kot93'} in a 
generelized sense. As an object of the research in this work, we choose the 
Schr\"odinger equation.

\section{Formulation of the algorithm}
\label{2}
Let us begin with a definition of symmetry which we shall call generelized 
one, and which we put in the base of the search for new symmetries.
\par Let the equation be given in the space $R^n(x)$
\begin{equation}
\label{f1}
L\phi(x)=0{,}
\end{equation}
where $L$ is a linear operator.
\newtheorem{definition}{Definition}
\begin{sloppypar}
\begin{definition} By the symmetry of Eq. (1) we shall mean a set of 
operators $Q^{(p)}$, p=1,2,$\ldots$,n,$\ldots$ if any operator 
of the type $L^{(p-1)}Q^{(p)}$ transforms a solution $\phi(x)$ into
another solution $\phi'(x)=L^{(p-1)}Q^{(p)}\phi(x)$.
\end{definition}
\end{sloppypar}
Definition leads to the operators $Q^{(p)}$ satisfying commutational
relations of order p
\begin{equation}
\label{f2}
[L\ [L\ldots[L,Q^{(p)}] \ldots]\ ]_{(p-fold)}\phi(x)=0
\end{equation}
\par  The generelized Definition:
\begin{itemize}
\item 
includes the understanding of symmetry also in the case that Eq. (2) is
fulfiled on a set of arbitrary functions, i.e. that
$[L\ [L\ldots[ L, Q^{(p)} ]\ldots]]_{(p-fold)}=0$ \cite {Kot83};
\item
contains the standard understanding of symmetry
$[L,Q^{(1)}] \phi(x)=0$ with p=1 \cite{Fus83}, \cite{Mal79};
\item
includes understanding of symmetry in quantum mechanics sense
$[L,Q^{(1)}]=0$ \cite{Pet67};
\item
differs from the standard one in accordance with which 
the framework of the latter by the operators of symmetry we must mean
not the  operators $Q^{(p)}$,  but the operators
$X^{(1)}=L^{(p-1)}Q^{(p)}$ \cite{Lez86}.
\end{itemize}
The question is how practically to find them. In the present work it
is made by analogy with the modifed Lie algorithm \cite {Nie72},
\cite {Fus90}. Below we shall consider the case with p=2.
\par Let us introduce a set of operators
\begin{equation}
\label{f4}
Q^{(2)}=\xi_2^a(x)\partial_a+\eta_2(x){,}
\end{equation}
which have the following commutation properties
\begin{equation}
\label{f6}
[L\ [L,Q^{(2)}]\ ]=\zeta_2(x)L
\end{equation}
The given expression is operator's version of the generelized Definition 
of symmetry with p=2. Here $\partial_a=\partial/\partial{x^a}$;
$a=0,1,\ldots{n-1}$; $\xi_{2}^{a}(x)$, $\eta_{2}(x)$, $\zeta_{2}(x)$ \ are 
unknown functions; the summation is carried out over a twice repeating index.
The unknown functions may be found by equaling the coefficients at
identical derivatives in the left and in the right sides of the ratios 
(\ref{f6}) and by integrating the resulting set of differential 
equations. This set we shall call the determining one by analogy with 
\cite{Fus90}. 
\par    After integrating the general form of the operator $Q^{(2)}$ may be
written as a linear combination of the basic elements 
$Q_\mu^{(2)}$ and $Q_\nu^{(2)}$, on which, by analogy with
\cite{Fus90}, we impose the condition of belonging to Lie algebras:
\begin{equation}
\label{f8}
[Q_\mu^{(2)},Q_\nu^{(2)}]=C_{\mu\nu\sigma}Q_\sigma^{(2)}
\end{equation}
Here \ $C_{\mu\nu\sigma}$ \ are the structural constants; operators  
$Q_\sigma^{(2)}$ belong to the \ {sets} of operators $Q^{(2)}$.
\par  By integrating the Lie equations we shall transfer from the Lie
algebras to the Lie groups
\begin{equation}
\label{f9}
dx^a{'}/d\theta=\xi^a{(x')}, \ x^a{'}_{(\theta=0)}=x^a,
\end{equation}
where $a=0,1,\ldots{n-1}$; \ $\theta$ \ is a group parameter \cite{Ibr83}, 
\cite{Fus90}.
\par   For the law of transforming the field $\phi (x)$ to be found,
instead of integrating the corresponding Lie equations \cite{Ibr83},
\cite{Fus90}
\begin{equation}
\label{fa}
d\phi'{(x')}/d\theta=\eta{(x')}\phi'{(x')}, \
\phi'{(x')}_{(\theta=0)}=\phi{(x)}
\end{equation}
we shall use the approach \cite{Kot86}, \cite{Kot91} that we shall 
illustrate by example of one-component field. We shall introduce a 
weight function $\Phi(x)$ into the field transformation law so that
\begin{equation}
\label{f10}
\phi'(x')=\Phi(x)\phi(x)
\end{equation}                                              
We shall choose the function $\Phi(x)$ so that Eq. (1) should transform
into itself
\begin{equation}
\label{f11}
\begin{array}{c}
\displaystyle L'\phi'(x')=0\to{x'=x'(x)}\to L\phi(x)=0{,}\\
\displaystyle L'=L
\end{array}
\end{equation}
due to the additional condition, namely, compatibility of the set of the
engaging equations
\begin{equation}
\label{f12}
\begin{array}{c}
\displaystyle A\Phi(x)\phi(x)=0{,}\\
\displaystyle L\phi(x)=0
\end{array}
\end{equation}
The former is obtained by replacing the variables in the initial
equation $L'\phi'(x')=0$.
If $A=L$, the symmetry will be called the classical or standard 
one; and the symmetry will be called the symmetry in generalized sense or 
the non-standard one if $A\not=L$. 
By solving Set (\ref{f12}) the weight function $\Phi(x)$ may be put in 
conformity to each field function $\phi(x)$ for ensuring the transition 
(\ref{f11}).
\par  Instead of solving Set (\ref{f12}) the weight function may be found on 
the base of the symmetry approach. As  $\phi'(x')=Q^{(1)'} \phi(x')$ 
is a solution too, and $\phi'(x')= \Phi(x)\phi(x)$ we have \cite{Kot86}, 
\cite{Kot91}:
\begin{equation}
\label{f13}
\Phi(x)={\phi'(x'\to x)\over\phi(x)}\in\{{\phi(x'\to x)\over\phi(x)};
{1\over\phi(x)};
{Q_\alpha^{(1)'}\phi(x'\to x)\over\phi(x)};{[L',Q_\mu^{(2)'}]
\phi(x'\to x)\over
\phi(x)};\ldots\}
\end{equation}
Here the dots correspond to a consecutive action of the operators 
$Q_\alpha^{(1)'}$ and $[L',Q_\mu^{(2)'}]$ on a solution $\phi(x')$.
Thus, for the function $\Phi(x)$ to be found it is necessary to turn to the 
unprimed variables in the primed solution $\phi'(x')$ and to divide
the result available by the unprimed solution  $\phi(x)$ \cite{Kot86}.
\par  After finding the weight functions $\Phi(x)$ the task on the
symmetry of Eq. (1) for one-component field may be thought as completed in 
the definite sense, namely: the set of the operators of symmetry and the
appropriate Lie  lgebra are indicated for p=2; the group of symmetry is
restored by the given algebra; the transformational properties of the field
$\phi(x)$ are determined with the help of the weight functions.
\par The symmetries found in such a way are distinguished by two
symptoms from the ones found in accordance with the standard Lie algorithm in
the classical or the modified versions: we use the commutational relations
(\ref{f6}) of higher order than unit and the non-Lie condition of symmetry
(\ref{f12}). The point is that the equation $A\Phi\phi(x)=0$ beloning to Set 
(\ref{f12}) is not Lie invariant because $A\neq L$ in general case.
However at the set of the solutions \{$\Phi (x)\phi (x), \ \phi (x)$\}
equations (\ref{f12}) become compartible and it follows
from $L'\phi'(x')=0$  that $L\phi(x)=0$. It is the non-Lie condition of
symmetry (\ref{f12}) that is the reason for using the relation (\ref{f10})
instead of integrating the corresponding Lie equations (\ref{fa}). 
\par  Taking it into account , we note that stated algorithm allows
generalization for the case of multicomponent field and symmetries
of higher order than p=2 and consider the particular examples.

\section{Application of the algorithm to the Schr\"odinger equation}
\label{3}
Let t designate the time; x, y, z  are the space variables; m is the mass 
of a particle, $\phi(x)$ is the wave function. Then \cite{Dav73}
\begin{equation}
\label{f14}
L_s\phi(x)={(i\hbar\partial_t+{\hbar^2\over2m}\triangle)}\phi(x)=0
\end{equation}
We shall consider the given equation within the framework of the stated 
algorithm.

\subsection{The symmetry of the type p=1. The Galilei symmetry of the
Schr\"odinger equation}
\label{3.1}
\par  The symmetry properties of Eq. (\ref{f14}) with p=1 were investigated  
by Niderer \cite{Nie72}, Fushchich and Nikitin \cite{Fus90}, Hagen
\cite{Hag72}. It has been established that the invariance algebra of Eq.
(\ref{f14}) is the Lie algebra of the Schr\"odinger group $Sch_{13}$. The
algebra generators commute on solutions with the operator $L_s$ from Eq.
(\ref{f14}) and belong to the type of the operators $Q^{(1)}$ with p=1,
for example
\begin{equation}
\label{f15}
[L_s,H_1]\phi(x)=0
\end{equation}
Here $H_1=it\partial_x+(m/\hbar)x$ is the generator of the Galilei
transformations
\begin{equation}
\label{f16}
x'=x-Vt{,}\ y'=y{,}\ z'=z{,}\ t'=t{,}
\end{equation}
V is the velocity of the inertial reference $K'$ relative to $K$.
By means of Eq. (\ref{f16}) we may formulate the set of equations for
finding the weight functions $\Phi(x)$:
\begin{equation}
\label{f17}
\begin{array}{c}
\displaystyle
(L_s+i\hbar{V}\partial_x)\Phi(x)\phi(x)=0{;}\\
\displaystyle
L_s\phi(x)=0
\end{array}
\end{equation}
Putting the free Schr\"odinger equation solution \cite{Dav73}
\begin{equation}
\label{f18}
\phi(x)=exp\lbrack-{i\over\hbar}({\rm E}t-\bf x.p)\rbrack
\end{equation}
into the first equation of the set (\ref{f17}) or into the formula
(\ref{f13}) we have:
\begin{equation}
\label{f19}
\Phi(x)={\phi(x'\to
x)\over\phi(x)}=exp\lbrack-{i\over\hbar}(-{\sl E}t+x{\sl P})\rbrack
\end{equation}
Here ${\rm E}=m{\bf v}^2/2$; ${\bf p}=m{\bf v}$; ${\sl E}=mV^2/2$;
${\sl P}=mV$; ${\bf v}$ is the speed of a particle. The function (\ref{f19})
coincides with the one obtained as a result of integrating the Lie equations
\cite{Nie72}, \cite{Fus90}, \cite{Hag72}.
\par 
\begin{sloppypar}
Thus, as to the symmetry of the type p=1, the stated algorithm has resulted
in the known conclusion: the Galilei transformations (\ref{f16}) are the 
symmetry transformations of the Schr\"odinger equation if the latter are 
accompanied by the transformation of the wave function (\ref{f18}) following 
the rule
\end{sloppypar}
\begin{equation}
\label{f20}
\phi'(x')=\Phi(x)\phi(x)=
{exp\lbrack-{i\over\hbar}(-{\sl E}t+x{\sl P})\rbrack}\phi(x)
\end{equation}
\noindent Similarly, the weight functions may be also found for the other
space-time transformations which fall into the Schr\"odinger group. As a
consequence of Set (\ref{f17}), Galilei symmetry of the Schr\"odinger
equation is the non-standard one in sense of the terminology from Section 2.

\subsection{The symmetry of the type p=2. Relativistic symmetry of 
the Schr\"odinger equation}
\label{3.2}
\par The evidence for possible existence of the relativistic 
symmetry of the Schr\"odinger equation has already been obtained by Malkin 
and Man'ko \cite{Mal79}, Fushchich and Segeda \cite{Fus77}. In particular, 
Œ lkin and Œ n'ko showed that the Schr\"odinger equation with the Coulomb 
potential if considered in the momentum p-space could be reduced to the free
D'Alembert equation in the special basis $q=f(p)$. Hence, it
is conform-invariant in the q-space \cite{Mal79}.
Fushchich and Segeda found the symmetry of the free Schr\"odinger equation 
with respect to the non-Lie, integro-differential operators which form the
representation of the Lie algebra of the Lorentz group \cite{Fus77}.  
However below we shall not consider these works in details as the
present paper is devoted to the investigation of the Lie invariance algebras
in the 4-dimentional real space-time.
\par It is known that the generators of the Lorentz transformations 
$M_{0k}=x_0\partial_k-x_k\partial_0$ \ ($x^0=c{t}$;\ $x_k=-x^k$;\ 
$x^k=x,y,z$) and the operator $L_s$ of the Schr\"odinger \ {equation}
satisfy the commutational relations of second order \cite{Kot83'}:
\begin{equation}
\label{f21}
[L_s\ [L_s,M_{0k}]\ ]=0
\end{equation}
Here $[L_s,M_{0k}]=\hbar(i{c}\partial_k+\hbar{\partial_t}{\partial_k}
/m{c})$. Hence, the generators $M_{0k}$ are the symmetry operators of
the Schr\"odinger equation in accordance with Definition 1 in Minkowski
space if p=2 \cite{Kot83'}.
Therefore, besides the Galilei interpretation the Schr\"odinger equation 
admits also the relativistic one. Accordingly, we assume that the mass
in Eq. (\ref{f14}) depends on the speed and follows the relativistic law  
$m=m_0/\sqrt{1-\beta^2}$, where $\beta=v/c$; $v$ is the velocity of a 
particle; $c$ is the speed of light; $m_0$ is the rest mass. Then we have:
\begin{equation}
\label{f22}
L_s^r\phi{(x)}=(i\hbar\partial_t+{\hbar^2\sqrt{1-\beta^2}\over2{m_0}}\triangle)
\phi(x)=(i\hbar\partial_t+{c^2\hbar^2\over2{W}}\triangle)\phi(x)=0
\end{equation}
We shall call this equation the relativistic Schr\"odinger equation and
give its solutions
\begin{equation}
\label{f23}
\phi_1^r{(x)}={exp\lbrack-{i\over\hbar}{({\beta^2\over2}W{t}-\rm \bf {P.x})}
\rbrack}=
exp\lbrack-i{mv^2\over2\hbar}(t-{{\bf {n.x}}\over{v}/2})\rbrack
\end{equation}
\begin{equation}
\label{f24}
\phi_2^r{(x)}={exp\lbrack-{i\over\hbar}(Wt-{\sqrt2\over\beta}
\rm {\bf P.x})\rbrack}=
exp\lbrack-i{mc^2\over\hbar}(t-{{\bf {n.x}}\over{c}/\sqrt2})\rbrack
\end{equation}
Here $W=mc^2$ \ and ${\rm {\bf P}}=m\bf v$ \ are the relativistic
energy and momentum of a particle; ${\bf n}={\bf v}/v$ is the guiding
vector of the velocity  $\bf  v$.
In the non-relativistic approach the solution $\phi_1^r{(x)}$ is reduced 
to the solution (\ref{f18})
\begin{equation}
\label{f25}
\phi_1^r{(x,\beta\ll 1)}\cong \phi_1^{nr}{(x)}=exp\lbrack-i{m_0{v^2}
\over2\hbar}(t-{{\bf {n.x}}\over v/2})\rbrack
\end{equation}
The solution $\phi_2^r{(x)}$ takes the form
\begin{equation}
\label{f26}
\phi_2^r{(x,\beta\ll 1)}\cong \phi_2^{nr}{(x)}= exp\lbrack-i{m_0{c^2}
\over\hbar}(t-{{\bf {x.n}}\over c/\sqrt2})\rbrack
\end{equation}
It is the new solution of the Schr\"odinger equation (\ref{f14}). It contains
the information on the value of the particle rest mass and
on the direction of its movement. Corresponding to this solution the wave 
(\ref{f26}) may be propagated with far higher speed $c/\sqrt2$ than the 
velocity of a particle $v\ll c$ and than the speed of the wave propagation 
$v/$2 from the solution (\ref{f25}).
\par Let us find the set of equations for the weight function $\Phi{(x)}$
from Eq. (\ref{f10}). We shall introduce the Lorentz transformations
\begin{equation}
\label{f27}
x'={{x-Vt}\over\sqrt{1-V^2/c^2}}{,}\ y'=y{,}\ z'=z{,}\ t'={{t-xV/c^2}
\over\sqrt{1-V^2/c^2}}
\end{equation}
By replacing the variables in Eq. (\ref{f22}) and taking into account the 
transformational properties of the mass, the energy, the momentum, as well
as the interrelations between the private derivatives we find
\begin{equation}
\label{f28}
\begin{array}{c}
\vspace{2mm}
\displaystyle
\lbrack i\hbar(\partial_t+V\partial_x)+ {c^2\hbar^2(1-V^2/c^2)\over
2W(1-{\bf
V.v}/c^2)}\lbrack{(\partial_x+V\partial_t/c^2)^{2}\over(1-V^2/c^2)}+
\partial_y{_y}+\partial_z{_z}\rbrack\rbrack\Phi{(x)}\phi{(x)}=0{;}\\
\displaystyle
L_s^r\phi{(x)}=0
\end{array}
\end{equation}
For the wave functions $\phi_1^r{(x)}$ and $\phi_2^r{(x)}$ from the
formulas (\ref{f23}) and (\ref{f24}) the solutions of the Set
(\ref{f28}) take the following forms:
\begin{equation}
\label{f29}
\begin{array}{c}
\displaystyle \Phi_1=exp\{-{i\over2\hbar(1-V^2/c^2)}\{\lbrack(\beta{'}^2-
2{V^2\over{c^2}}-\beta^2(1-{V^2\over{c^2}})){W}-\\
\displaystyle V(\beta{'}^2-2)P_x\rbrack t-
(\beta{'}^2-2)({V{W}\over{c^2}}-{V^2P_x\over{c^2}})x\}\}{;}
\end{array}
\end{equation}
\begin{equation}
\label{f30}
\begin{array}{c}
\displaystyle \Phi_2=exp\{-{i\over\hbar(1-V^2/c^2)}\{(1-{\sqrt2\over
\beta'})({V^2W\over{c^2}}-VP_x)t+
\lbrack{V^2\over{c^2}}(1-{\sqrt2\over\beta'})+\\
\displaystyle \sqrt2({1\over\beta}-{1\over\beta'})\rbrack P_x{x}-
(1-{\sqrt2\over\beta'}){W{V}{x}\over{c^2}}+
(1-{V^2\over{c^2}})
\sqrt2({1\over\beta}-{1\over\beta'})(P_y{y}+P_z{z})\}\}
\end{array}
\end{equation}
Here $\beta{'}^2=\lbrack V^2(1-\beta^2)/c^2+\beta^2-2V\beta_x/c+
V^2\beta_x^2/{c^2}\rbrack/(1-V\beta_x/c)^2$; $V$ is the velocity of
frame $K'$ relative to $K$; $c$ is the speed of light; $\beta=v/c$;
$\beta_x=v_x/c$.
\par    In the non-relativistic approximation Set (\ref{f28})  
is turned into  Set (\ref{f17}), and the function 
$\Phi_1{(x,\beta\ll 1)}$ from Eq. (\ref{f29}) coincides with the function 
(\ref{f19}). The non-relativistic limit of the function $\Phi_2(x)$ is
\begin{equation}
\label{f31}
\begin{array}{c}
\displaystyle \Phi_2{(\beta\ll 1)} \cong exp\{-i\sqrt2{m_0{c}V(v_x-V)
\over\hbar v'}\{t-\lbrack n_x(v-v')-V\rbrack{x\over V{(v_x-V)}}-\\
\displaystyle n_y(v-v'){y\over V{(v_x-V}})-n_z(v-v'){z\over V{(v_x-V)}}\}{,}
\end{array}
\end{equation}
where $v'={(v^2-2V{v_x}+V^2)}^{1/2}$.
\par  In another case, when $\beta=1$, the relativistic Schr\"odinger
equation describes the propagation of the fields of wave functions
corresponding to the movement of a massless particle with the velocity
$c$, the momentum ${\bf P}={\bf n}P$ and the energy $W=cP$:
\begin{equation}
\label{f32}
(i\hbar\partial_t+{c^2\hbar^2\over2{W}}\triangle)\phi(x)=
(i\hbar\partial_t+{c\hbar^2\over2{P}}\triangle)\phi(x)=0
\end{equation}
The solutions of this equation may be derived from the formulae (\ref{f23})
and (\ref{f24}) if $\beta=1$.
\begin{equation}
\label{f33}
\phi_1^r(x,\beta=1)=exp\lbrack-i{cP\over2\hbar}(t \ -{{\bf n.x}\over{c}/2})
\rbrack{;}
\end{equation}
\begin{equation}
\label{f34}
\phi_2^r(x,\beta=1)=exp\lbrack-i{cP\over\hbar}(t \ -{{\bf n.x}\over{c}/
\sqrt2})\rbrack
\end{equation}
Here ${\bf n}={\bf c}/c$, the phase velocity of propagation for the first 
wave is equal to $c/2$ and $c/\sqrt2$ for the second one.
In both cases these velocities are less than the speed of light $c$. As a 
result two wave functions with the different speeds of propagation and the
different frequencies $\omega_1=cP/2\hbar$ and $\omega_2=cP/\hbar$ 
correspond to the same value of the momentum and the energy of a
particle.\footnote{Taking into account the
sense of values $\omega_{1}$ and $\omega_{2}$, we may interprete the functions
(\ref{f33}) and (\ref{f34}) as the solutions of D'Alembert equation in 
Minkowski space $M(wt,{\bf x})$ when velocities of fields propagation are
$w_{1}=c/2$ and $w_{2}=c/\sqrt2$ respectively. Because of the relativistic
symmetry of D'Alembert equation, this possibility is additional illustration 
to the relativistiic symmetry of Schr\"odinger equation too. A
representation of the Lie algebra of Poincare' group may be realized on the 
solutions of the free Schr\"odinger equation.}
The corresponding weight functions are:
\begin{equation}
\label{f35}
\Phi_1{(x,\beta=1)}=exp\{-{iVP\over2\hbar(1-V^2/c^2)}\lbrack
(n_x-{V\over c}){t}+(1-n_x{V\over c}){x\over c}\rbrack\}{;}
\end{equation}
\begin{equation}
\label{f36}
\Phi_2{(x,\beta=1)}=exp\{-{i(\sqrt2-1){V}P\over\hbar(1-V^2/c^2)}
\lbrack(n_x-{V\over c}){t}+(1-n_x{V\over c}){x\over c}\rbrack\}
\end{equation}
\par    Thus, the relativistic symmetry of the Schr\"odinger equation
contribute to the information on the properties and the solutions of
this equation. In particular, it enables the equation to be extended to the 
area of the relativistic values of mass, momentum and energy of the particle
under study.
It results in the existence of two different solutions, corresponding
to the same particle. The phase velocity of propagation of the first
wave is equal $v/2$; the phase velocity of the second wave is equal
$c/\sqrt2$, where $v$ is the particle velocity and $c$ is the speed
of light. In the non-relativistic approximation the first solution 
(\ref{f25}) has the known form (\ref{f18}). The second solution 
(\ref{f26}) contains the information on the particle rest mass and on 
the direction of its movement.
The phase velocity of propagation of the second wave much more exceeds both
the non-relativistic speed of a particle $v\ll c$ and the phase velocity
of propagation of the wave $v/2$ corresponding to the first solution.
\par  In the case of massless particles the propagation velocities of 
the fields $\phi_{1}^{r}(x)$ and $\phi_{2}^{r}(x)$ are equal to $c/2$
and $c/\sqrt2$ respectively. The fields lag behind the particle moving with
the speed of light. It is similar to propagation of Cherenkov radiation 
or sound from a source moving with a supersonic velocity.
The transformation law of the fields of this particle may be written as
follows:
\begin{equation}
\label{fad}
\Psi_{P,c/2,c/\sqrt2}'=\Phi \Psi_{P,c/2,c/\sqrt2} \to
\left( \begin{array}{c}
\phi_{1}^{r}{'}\\ 
\phi_{2}^{r}{'}
\end{array} \right )=
\left( \begin{array}{cc}
\Phi_{11} & 0\\
0 & \Phi_{22}
\end{array} \right )
\left( \begin{array}{c}
\phi_{1}^{r}\\
\phi_{2}^{r}
\end{array} \right )=
\frac{1}{2} \left( \begin{array}{cc}
\Phi_{11} & \Phi_{12}\\
\Phi_{21} & \Phi_{22}
\end{array} \right ) 
\left( \begin{array}{c}
\phi_{1}^{r}\\
\phi_{2}^{r}
\end{array} \right ),
\end{equation}
\noindent where 
$\Phi_{12}=\Phi_{11}\phi_{1}^{r}/\phi_{2}^{r}$,
$\Phi_{21}=\Phi_{22}\phi_{2}^{r}/\phi_{1}^{r}$;
$\Phi_{11}$ and $\Phi_{22}$ are the weight functions (\ref{f35}) and 
(\ref{f36}) respectively.
\par  Thus, the relativistic symmetry of the Schr\"odinger equation
allows the possibility of existence of massless fields propagating with
the speeds smaller than the speed of light. It permits the possibility
of existence of some hypothetical massless particles with the set
of continuos parameters (momentum $P$, energy $W$) and discrets ones ($c$, 
$c/2$, $c/\sqrt2$).
\par    Within the framework of the accepted terminology the relativistic
symmetry of the Schr\"odinger equation is the symmetry in generelized sense.

\subsection{Maximal linear group symmetry of the Schr\"odinger
equation with p=2}
\label{3.3}
The maximal group of symmetry (the algebra of invariance) of an equation
$L\phi{(x)}=0$ means the symmetry group (the algebra of invariance)
of the maximal dimension, permitted by the given equation in Lie sense.
In the present work the concept of maximality is not
unambiguous as it must be conformed with the value of the number p
in the commutational ratio (\ref{f2}). It is necessary to distinguish between
the maximal dimensions of group or algebra with p=1, p=2 and etc.
\par    For determining the dimensions of invariance algebras of the
Schr\"odinger equation we shall start, by analogy with the modified Lie
algorithm \cite{Nie72}, \cite{Fus90} from the set of the determining 
differential equations corresponding to the conditions (\ref{f6}).

For example, first we will write the equations system in the case of p=1.
By equaling the coefficients at identical derivatives in the left and the 
right sides of the ratios $[L_{s},Q^{(1)}]=\zeta_{1} (x)L_{s}$, where
$Q^{(1)}=\xi_{1}^{a}(x)\partial_{a}+\eta_{1}(x)$, we have:
\begin{equation}
\label{f}
\begin{array}{llll}
\vspace{1mm}
\displaystyle
\partial_{1}\xi_{1}^{1}= & 
\partial_{2}\xi_{1}^{2}= & 
\partial_{3}\xi_{1}^{3}{;} &
{} \\
\vspace{1mm}
\displaystyle
\partial_{1}\xi_{1}^{2}+\partial_{2}\xi_{1}^{1}= &
\partial_{2}\xi_{1}^{3}+\partial_{3}\xi_{1}^{2}= &
\partial_{3}\xi_{1}^{1}+\partial_{1}\xi_{1}^{3}{;} &
{} \\ 
\vspace{3mm}
\displaystyle
\partial_{11}\xi_{1}^{0}=0{;} & 
{} & 
{} &
{} \\
\vspace{3mm}
\displaystyle 
L_{s}\xi_{1}^{0}=i\hbar\zeta_1{;} &
L_{s}\xi_{1}^{k}=-i(\hbar^{2}/m)\partial_{k}\eta_{1}, & 
k=1,2,3{;} & 
{} \\ 
\displaystyle 
L_{s}\eta_{1}=0 & 
{} &
{} &
{} \cr
\end{array}
\end{equation}
\noindent Providing $\hbar=c=1$, the system may be transformed into the
system from \cite{Fus90}. The system solutions allow the general expression
for symmetry operator $Q^{(1)}$ to be writted as the linear combination of
operators \cite{Fus90}:
\begin{equation}
\label{ff}
Q^{(1)}=a^{0}p_0 + a^{k}p_k + b^{k}J_k + d^{k}H_k + eM + gD + fK,
\end{equation} 
\noindent where $a^0, a^k, b^k, d^k, e, g, f$ are the arbitrary numbers;
$x^{0}=t$,\ $x^{k}=-x_{k}=(x,y,z)$, k=1,2,3; $L_{s}=i\hbar \partial_{0}+
(\hbar^{2}/2m)\Delta$;\ $p_0=i\partial_{t}$; \  $p_k=-i\partial_{k}$; \
$J_k=({\bf x}$x${\bf p})_{k}$; \  $H_k=-tp_{k}+Mx_{k}$; \ $M=m/\hbar$; \ 
$D=2tp_{0}-{\bf xp}+3i/2$; \  $K=t^{2}-tD-M{\bf x}^{2}/2$ are the generators
of Lie algebra of Schr\"odinger group $Sch_{13}$ \cite{Nie72}, \cite{Fus90},
\cite{Hag72}. From this it follows, that the maximal invariance algebra of 
Schr\"odinger equation with p=1 is the algebra Lie of the Schr\"odinger group
$Sch_{13}$ \cite{Nie72}, \cite{Fus90}, \cite{Hag72}.
\par In the case of the type p=2 symmetry, when $[L_{s} \ [L_{s},Q^{(2)}]]=
\zeta_{2} (x)L_{s}$ and $Q^{(2)}$ is the operator (\ref{f4}), the analogue 
of Set (\ref{f}) takes the form \cite{Kot94}:
\begin{equation}
\label{fff}
\begin{array}{llll}
\vspace{1mm}
\displaystyle
\partial_{11}\xi_{2}^{1}=0{;} & 
\partial_{22}\xi_{2}^{1}=-2\partial_{12}\xi_{2}^{2}{;} &
\partial_{33}\xi_{2}^{1}=-2\partial_{13}\xi_{2}^{3}{;} & 
{} \\
\vspace{1mm}
\displaystyle
\partial_{11}\xi_{2}^{2}=-2\partial_{12}\xi_{2}^{1}{;} & 
\partial_{22}\xi_{2}^{2}=0{;} &
\partial_{33}\xi_{2}^{2}=-2\partial_{23}\xi_{2}^{3}{;} & 
{} \\
\vspace{1mm}
\displaystyle
\partial_{11}\xi_{2}^{3}=-2\partial_{13}\xi_{2}^{1}{;} &
\partial_{22}\xi_{2}^{3}=-2\partial_{23}\xi_{2}^{2}{;} &
\partial_{33}\xi_{2}^{3}=0{;} & 
{} \\
\vspace{1mm}
\displaystyle
\partial_{01}\xi_{2}^{1}= &
\partial_{02}\xi_{2}^{2}= &
\partial_{03}\xi_{2}^{3}{;} & 
{} \\
\vspace{3mm}
\displaystyle
\partial_{12}\xi_{2}^{3} + &
\partial_{23}\xi_{2}^{1} + &
\partial_{31}\xi_{2}^{2} =0{;} & 
{} \\
\vspace{1mm}
\displaystyle
L_{s}\partial_{0}\xi_{2}^{0}= &
-(\hbar^{2}/2m)\Delta\partial_{0}\xi_{2}^{0} &
+(i\hbar^{3}/4m^{2})\Delta^{2}\xi_{2}^{0} + \zeta_{2} &
{} \\
\vspace{1mm}
\displaystyle
L_{s}\partial_{k}\xi_{2}^{0}= &
+(\hbar^{2}/2m)\Delta\partial_{k}\xi_{2}^{0}{;} &
{} &
{} \\
\vspace{1mm}
\displaystyle
L_{s}\partial_{0}\xi_{2}^{k}= &
-(\hbar^{2}/2m)(\Delta\partial_{0}\xi_{2}^{k}+4\partial_{k0}\eta_{2}) &
+(i\hbar^{3}/4m^{2})(\Delta^{2}\xi_{2}^{k}+4\Delta\partial_{k}\eta_{2}){;} &
{} \\
\vspace{1mm}
\displaystyle
L_{s}\partial_{1}\xi_{2}^{1}= &
+\zeta_{2}/4 &
-(\hbar^{2}/2m)\partial_{11}\eta_{2} &
{} \\
\vspace{3mm}
\displaystyle
L_{s}(\partial_{j}\xi_{2}^{k}+\partial_{k}\xi_{2}^{j})= &
-(\hbar^{2}/2m)\partial_{jk}\eta_{2} &
j\neq k &
{} \\
\vspace{1mm}
\displaystyle
\partial_{11}\eta_{2}= &
\partial_{22}\eta_{2}= &
\partial_{33}\eta_{2}{;} &
{} \\
L_{s}\partial_{0}\eta_{2}= &
+(i\hbar^{3}/4m^{2})\Delta^{2}\eta_{2} &
{} &
{} \cr
\end{array}
\end{equation}
\noindent Set (\ref{fff}) has the solution permitting the symmetry operator
$Q^{(2)}$ to be writted as:
\begin{equation}
\label{ffff}
Q^{(2)}=a^{0}P_{0} + a^{k}P_{k} + b^{0k}G_{0k} + b^{lm}G_{lm} + \ldots,
\end{equation}
\noindent where $a^{0}, a^{k}, b^{0k}, b^{lm}$ are the arbitrary numbers;
$k,l,m=1,2,3$; the summation is absent over a twice repeating index;
$P_{0}=\partial_{0}$, $P_{k}=\partial_{k}$, $G_{0k}=x_{0}\partial_{k}$,
$G_{lm}=x_{l}\partial_{m}$ are the generators of the 20-dimensional Lie 
algebra
\begin{equation}
\label{f37}
[P_a, P_b] = 0{;}\  [P_a, G_{bc}] = \delta_{ab} P_c{;}\  
[G_{ab}, G_{cd}] = \delta_{bc}G_{ad} - \delta_{ad}G_{cb}
\end{equation}
Here $\delta_{ab}=0$ at $a\ne b$, $\delta_{aa}=1$, $a,b=0,1,2,3$.
The algebra genertators satisfy the commutational ratios
\begin{equation}
\label{f38}
[L_{s} \ [L_s, P_a]]=0{;} \  [L_{s} \ [L_s, G_{ab}]]=0
\end{equation}
where $[L_{s},P_{a}]=0$, \ $[L_s, G_{ab}]=i\hbar \delta_{0a} \partial_b + 
(\hbar^2/2m) \delta_{ak}\partial_{kb}$. Hence, the algebra (\ref{f37}) 
is the algebra of invariance of the Schr\"odinger equation when p=2. 
It induces the 20-dimensional group IGL(4,R) of the non-uniform linear 
transformations of space-time variables in the space $R^4{(x)}$:
\begin{equation}
\label{f39}
x^a{'}=A_b^a{x^b} + A^a
\end{equation}
Here all the sixteen matrix elements $A_b^a$ are different in general.
As far as more total linear transformations do not exist in the space 
$R^4{(x)}$, the group IGL(4,R) forms the maximal linear
group of symmetry of the Schr\"odinger equation when p=2. All the other
space-time transformations, corresponding to the condition (\ref{f6}), 
are nonlinear ones. As an example, we shall show the transformations 
induced by the operators  
\begin{equation}
\label{f40}
Q_1^{(2)}=2x^0{^2}\partial_0 + x^0{x^k}\partial_k{;}
\end{equation}
\begin{equation}
\label{f41}
Q_{jk}^{(2)}=x^0({x^j}\partial_k - x^k\partial_j),
\end{equation}
where the functions $\xi_{2}^{a}$ and $\eta_{2}$ are the solutions 
of Set (\ref{fff}). The operators (\ref{f40}) and (\ref{f41}) satisfy the 
relations $[L_{s} [L_{s},Q_{1}^{(2)}]]=4i\hbar L_{s}$, \
$[L_{s} [L_{s},Q_{jk}^{(2)}]]=0$.
We have by integrating the Lie equations (\ref{f9})
($dx^{0}{'}/d\theta$ = $2x^{0}{'}^{2} \to 1/2x^{0}{'}$ = $-\theta+1/2x^{0}$; 
$dx^{k}{'}/d\theta$ = $x^{0}{'}x^{k}{'} \to dx^{k}{'}/x^{k}{'}$ = 
$x^{0}d\theta/(1-2\theta x^{0})$;
$dx^{j}{'}/d\theta$ = $-x^{0}{'}x^{k}{'}$, $dx^{k}{'}/d\theta$ = 
$x^{0}{'}x^{j}{'} \to x^{j}{'}dx^{j}{'}$ = $-x^{k}{'}dx^{k}{'} 
\to x^{j}{'}^{2}+x^{k}{'}^{2}$ = ${x^{j}}^2+{x^{k}}^2$):
\begin{equation}
\label{f42}
Q_1^{(2)}: x^0{'}=\frac{x^0}{1 - 2\Theta_1 x^0}; \
{\bf x}{'}=\frac{\bf x}{\sqrt{1-2\Theta_1 x^0}}{;}
\end{equation}
\begin{equation}
\label{f43}
\begin{array}{l}
\vspace{2mm}
\displaystyle Q_{jk}^{(2)}: x^0{'}=x^0;\\ 
\vspace{2mm}
\displaystyle x^j{'}=x^j cos(\Theta_{jk} x^0)-x^k sin(\Theta_{jk} x^0);\\  
\vspace{2mm}
\displaystyle x^k{'}=x^j sin(\Theta_{jk} x^0)+x^k cos(\Theta_{jk} x^0)
\end{array}
\end{equation}
Here $\Theta_1$ and $\Theta_{jk}$ are the group parameters. As a result
a theorem takes place.
\newtheorem{theorem}{Theorem}
\begin{theorem}  The group IGL (4,R) is the maximal linear
group of symmetry of the free Schr\"odinger equation in the 4-dimensional
real space-time $R^4(x)$.
\end{theorem}
\par  Accordingly to Theorem the Galilei and relativistic symmetries are 
the natural properties of the Schr\"odinger equation, since the Galilei, 
Lorentz and Poincar\'e groups are the IGL(4,R) subgroups. The generators 
of these transformations may be constructed from the generators of the 
algebra (\ref{f37}) by their linear combinations. Since
$[\Box [\Box ,P_a]]=0$, \ $[\Box [\Box ,G_{ab}]]=0$, where  
$[\Box ,P_a]=0$, \ $[\Box ,G_{ab}]=
2(\delta_{0a}\partial_{0b}-\delta_{ka}\partial_{kb})$, this group is the 
maximal linear group not only for the Schr\"odinger equation, but and for 
D'Alembert and Maxwell equations too. Thus, the IGL(4,R) group is the maximal 
linear symmetry group of the type p=2 both in the quantum mechanics, and in 
the classical electrodynamics.

\subsection{The infinite algebra of invariance of the Schr\"odinger equation
with p $\to \infty$}
\label{3.4}

    Let us turn to the formulae (\ref{f10}) and (\ref{f13}) from the algorithm 
of section 2. It can be seen, that any equation $L\phi(x) = 0$ for 
one-component field and, in particular, the Schr\"odinger equation has the 
property of invariance under arbitrary reversible transformations $x'=x'(x)$, 
$x=x(x')$ due to the following rations
\begin{equation}
\label{f44}
\begin{array}{c}
\displaystyle
L'\phi'(x')=0\to {x'=x'(x),\ \phi'(x')=\Phi(x)\phi(x)}\to
A\Phi(x)\phi(x)=0{;}\\
\displaystyle
L\phi(x)=0,
\end{array}
\end{equation}
where the set of engaging equations $A\Phi(x)\phi(x)=0, \
L\phi(x)=0$ satisfies the condition of the compatibility, if the weight
function is choosen in the form (\ref{f13}): $\Phi(x)=\phi'(x')/\phi(x)$.
The concept of maximality of a symmetry group then becomes uncertain in view
of arbitrary of transformations x'=x'(x).
\par    In the case of the algebraic approach the given property of the 
Schr\"odinger equation results in that the symmetry operators of
this equations may be constructed from elements of the infinite set of
operators
\begin{equation}
\label{f45}
\partial_a;\  x_b\partial_c;\  x_d{x_g}\partial_h;\  x_s{x_p}{x_q}
\partial_t,\ldots
\end{equation}
Here  a,b,\ldots, q,t \ =\ 0,1,2,3. These operators induce the infinite
Lie algebra
\begin{equation}
\label{f46}
\begin{array}{c}
\vspace{1mm}
\displaystyle [\partial_a, \partial_b]=0{;}\\
\vspace{1mm}
\displaystyle [\partial_a, x_b \partial_c]=\delta_{ab} \partial_c{;}\\
\vspace{1mm}
\displaystyle [\partial_a, x_b{x_c} \partial_d]=\delta_{ab} {x_c}\partial_d +
\delta_{ac} {x_b}\partial_d{;}\\
\vspace{1mm}
\displaystyle [\partial_a, x_b{x_c}{x_d}\partial_g]=
\delta_{ab} x_c{x_d}\delta_g +
\delta_{ac} x_b{x_b}\partial_g + \delta_{ad} x_b{x_c}\partial_g{;}
\end{array}
\end{equation}
\begin{equation}
\label{f47}
\begin{array}{c}
\vspace{1mm}
\displaystyle [x_a \partial_b, x_c \partial_d]=\delta_{bc} {x_a}\partial_d -
\delta_{ad} {x_c}\partial_b{;}\\
\vspace{1mm}
\displaystyle [x_a \partial_b, x_c{x_d}\partial_g]=
\delta_{bc} x_a{x_d}\partial_g +
\delta_{bd} x_a{x_c}\partial_g - \delta_{ag} x_c{x_d}\partial_b{;}\\
\vspace{1mm}
\displaystyle [x_a\partial_b, x_c{x_d}{x_g}\partial_h]=
\delta_{bc} x_a{x_d}{x_g}\partial_h +
\delta_{bd} x_a{x_c}{x_g}\delta_h + \delta_{bg} x_a{x_c}{x_d}\partial_h -
\delta_{ah} x_c{x_d}{x_g}\partial_b{;}
\end{array}
\end{equation}
\begin{equation}
\label{f48}
\begin{array}{c}
\vspace{1mm}
\displaystyle [x_a{x_b}\partial_c, x_d{x_g}\partial_h]=
\delta_{cd} x_a{x_b}{x_g}
\partial_h + \delta_{cg} x_a{x_b}{x_d}\partial_h - \delta_{ah} x_b{x_d}
{x_g}\partial_c - \delta_{bh} x_a{x_d}{x_g}\partial_c{;}\\
\vspace{1mm}
\displaystyle [x_a{x_b}\partial_c, x_d{x_g}{x_h}\partial_f]=
\delta_{cd} x_a{x_b}{x_g}{x_h}\partial_f +  
\delta_{cg} x_a{x_b}{x_d}{x_h}\partial_f  +\\
\vspace{1mm}
\displaystyle \delta_{ch} x_a{x_b}{x_d}{x_g}\partial_f - 
\delta_{af}x_b{x_d}{x_g}{x_h}\partial_c -
{\delta_{bf} x_a{x_d}{x_g}{x_h}\partial_c}{;}
\end{array}
\end{equation}
\dotfill
\par Let us assume $a=0$  in the formula (\ref{f46}), and take into account
the commutational ratios
\begin{equation}
\label{f49}
\begin{array}{c}
\vspace{1mm}
\displaystyle [\triangle, \partial_a]=0{;}\\
\vspace{1mm}
\displaystyle [\triangle, x_a\partial_b]=2\delta_{ka}\partial_{kb}{;}\\
\vspace{1mm}
\displaystyle [\triangle, x_a{x_b}\partial_c]=
2\delta_{ka}\delta_{kb}\partial_c +
2\delta_{ka} x_b\partial_{kc} + 2\delta_{kb} x_a\partial_{kc}{;}\\
\vspace{1mm}
\displaystyle [\triangle, x_a{x_b}{x_c}\partial_d]=
2\delta_{ka}\delta_{kb}  x_c
\partial_d + 2\delta_{ka}\delta_{kc} x_b\partial_d + \\
\vspace{1mm}
\displaystyle 2\delta_{kb}\delta_{kc} x_a\partial_d +
2\delta_{ka} x_b{x_c}\partial_{kd} +
2\delta_{kb} x_a{x_c}\partial_{kd} + 2\delta_{kc} x_a{x_b}\partial_{kd}
\end{array}
\end{equation}
It is possible to show then, that the generators (\ref{f45}) of the Lie
algebra of (\ref{f46}) - (\ref{f48})
and the operator of the Schr\"odinger equation $L_s$ satisfy the
commutational ratios (\ref{f2}). We omit the details, whose example can
be the formula
\begin{equation}
\label{f50}
\begin{array}{c}
\vspace{1mm}
\displaystyle[L_s\ [{L_s}\ [{L_s}, x_a{x_b}{x_c}\partial_d]]]=-6i\hbar^3
\delta_{0a}\delta_{0b}\delta_{0c}\partial_d -\\
\vspace{1mm}
\displaystyle 6(\hbar^4/m)(\delta_{ka}\delta_{0b}\delta_{0c} +
\delta_{kb}\delta_{0c}\delta_{0a} +
\delta_{kc}\delta_{0a}\delta_{0b})\partial_{kd} +\\
\vspace{1mm}
\displaystyle 6i(\hbar^5/{m^2})(\delta_{ka}\delta_{jb}\delta_{0c} +
\delta_{kb}\delta_{jc}\delta_{0a} + \delta_{kc}\delta_{ja}\delta_{0b})
\partial_{jkd} + \\
\vspace{1mm}
\displaystyle 2(\hbar^6/{m^3})(\delta_{ka}\delta_{jb}\delta_{mc} +
\delta_{kb}\delta_{jc}\delta_{ma} + \delta_{kc}\delta_{ja}\delta_{mb})
\partial_{jkmd}{,}
\end{array}
\end{equation}
and demonstrate only the final result:
\begin{equation}
\label{f51}
\begin{array}{c}
\vspace{1mm}
\displaystyle [L_s, \partial_a]=0{;}\\
\vspace{1mm}
\displaystyle [L_s\ [L_s, x_a\partial_b]\ ]=0{;}\\
\vspace{1mm}
\displaystyle [L_s\ [L_s\ [L_s, x_a{x_b}\partial_c]\ ]\ ]=0{;}\\
\vspace{1mm}
[L_s\ [L_s\ [L_s\ [L_s, x_a{x_b}{x_c}\partial_d]\ ]\ ]\ ]=0
\end{array}
\end{equation}
\dotfill

\noindent Here j, k, m =1,2,3;  the other indexes are
0,1,2,3; the summation is carried out over a twice repeating index; for 
the formulae (\ref{f51}) to be valid it is necessary, that the order of the
commutational ratio would exceed the summary degree of space-time variables 
in generators of the Lie algebra by unit. A theorem results from it.
\begin{sloppypar}
\begin{theorem} The infinite Lie algebra of the operators (\ref{f45}) is the
algebra of invariance of the free Schr\"odinger equation  when 
p $\to \infty$.
\end{theorem}
\end{sloppypar}
\par    In addition, it can be seen, that this algebra is the invariance 
algebra for the D'Alembert equation too.

\section{The conclusion}
\label{5}
    The algorithm is proposed to study the symmetry properties
of theoretical and mathematical physics equations of the type $L\phi(x)=0$. 
The main distinctive features of this algorithm are:
\begin{itemize}
\item introduction of the operators of symmetry $Q^{(p)}$ satisfying
the commutational ratios of higher order than p=1:
$[L\ [L\ldots[L,Q^{(p)}]\ldots]_{(p-fold)}\phi(x)=0$;
\item introduction of some weight function $\Phi(x)$ which is not a
component of field into the law of field transformation
$\phi'(x')=\Phi(x) \phi(x)$;
\begin{sloppypar}
\item interpretation of the compatibility of the set of equations
$A\Phi(x)\phi(x)=0{,} \  L\phi(x)=0$,
where $A\Phi(x)\phi(x)=0$ is obtained by replacing the variables 
$x'=x'(x)$ in the initial equation $L'\phi'(x')=0$, as the condition   
of transformation into itself of the initial equation 
$L'\phi'(x')=0\to L\phi(x)=0, \  L'=L$.
\end{sloppypar}
\end{itemize}

\par The application of the algorithm to the Schr\"odinger equation
(earlier to the D'Alembert and Maxwell equations) has allowed us to
establish, that in addition to the known (standard, p=1) symmetry the
Schr\"odinger equation has the relativistic symmetry (the D'Alembert and
Maxwell equations have the Galilei symmetry) when p=2. This circumstance
permits the Schr\"odinger equation to be extended to the area of relativistic
movements with relativistic values of mass, energy and momentum of the
particle under study. The Galilei and the relativistic symmetries are the
particular realizations of the more general symmetry of equations with
respect to the 20-dimensional group of non-uniform transformations IGL(4,R)
in the real space-time $R^4{(x)}$ with p=2. When $p\to \infty$, the
Schr\"odinger and D'Alembert equations display the symmetry with respect to
the infinite Lie algebra which contains the Lie algebras of the groups
IGL(4,R), Galilei, Lorentz and Poincar\'e as the subalgebras.
\par The relativistic symmetry of the Schr\"odinger equation permits 
the possibility of existence of some hypothetical massless particle, moving
with the speed of light $c$ and characterized by the two wave fields
propagating behind of the particle with the speeds $c/2$ and $c/\sqrt2$.
\par    In summary, it is  possible to state that the concept of symmetry
is conventional. It depends on the definition of symmetry and the associated
algorithm. Dividing the equations into the relativistic and the
Galilei-invariant equations makes sense only in the case of the narrow
understanding of symmetry when p=1. In more general case, when  $p\geq 1$,
equations have cumulative symmetrical properties complying with the
principles of relativity in the relativistic, in the Galilei, as well as
in the other versions. In accordance with F. Klein's \cite{Kle85}, we may
state that the equation satisfies as many principles of relativity, as
groups of symmetry exist for this equation.

\section{Acknowledgments}
\label{6}
     The author is deeply grateful to Prof. V. I.  Man'ko and Prof.  V. I.
Fushchich  for discussing  various fragments of the present work and
valuable remarks.

\end{document}